# Closely piling up of multiple adhesive fronts in adhesive friction due to re-attachment

Puyu Cao[1], Meicheng Yao[1], Bin Chen[1,2*]

[1]Department of Engineering Mechanics, Zhejiang University, Hangzhou, China

[2]Key Laboratory of Soft Machines and Smart Devices of Zhejiang Province, Hangzhou, China

* To whom correspondence should be addressed: chenb6@zju.edu.cn

**Abstract**

As a fundamental force, friction exerts a profound influence on various aspects of our daily lives across multiple disciplines. To understand why adhesive friction is associated with the contact area, here we investigate the generic sliding of elastic solids adhered to a rigid surface by considering re-attachment/healing. We then reveal multiple adhesive fronts closely aligning along the interface with the number of these regions generally increasing with the contact area. These adhesive fronts exhibit rich dynamics and their accumulation along an interface can aid each other through re-attachment/healing in friction, apparently resulting in the increase in the calculated shear-off force with the contact area. Based on these findings, we propose a refined law of adhesive friction. Our analysis further suggests that accumulating adhesive fronts along the interface can trigger crack-like propagation of individual fronts at high velocities, which potentially bridges the gap between tribology and fracture mechanics. We also discuss the relevance of this work to earthquake mechanics, which might provide a unified framework that captures key aspects of fault behavior. We expect that this work can supply a fundamental understanding of healing-mediated interfacial phenomena in diverse systems spanning biology, geology, and engineering.

## Introduction

Friction, as a fundamental physical phenomenon, pervades nearly all aspects of the universe and plays diverse yet essential roles in daily life. Its manifestations span multiple scales: At the nanoscale, the spatula pads beneath gecko toes exploit van der Waals interactions to generate frictional adhesion, enabling them to effortlessly scale smooth surfaces [1,2]; At the microscale, migrating cells must overcome frictional forces at adhesion sites within their microenvironment [3–6]; At the macroscale, fault friction along tectonic plate boundaries governs stress accumulation and release, critically influencing earthquake dynamics [7–10]. Traditionally, friction obeys Amonton's law [11–14], where the frictional force scales linearly with the applied normal load. However, Bowden and Tabor [15] proposed that friction at microscopic level arises from the real contact area between weakly adhering surfaces. Their model aligns with Amonton's law if the real contact area is proportional to the normal load [12,16].

Experimental evidence supports Bowden and Tabor's proposition [15], demonstrating that frictional force correlates with the real contact area [17–25]. Interestingly, the proportionality constant between frictional force and the real contact area corresponds to the frictional shear strength of the interface [15], implying that adhesive friction is apparently not governed by an energy criterion in the classical fracture theory [26]. As an example, in the Kendall's peeling model of a thin film being perfectly adhered to a substrate [27], adhesive forces should concentrate within the size of the cohesive zone, a size inherently existing within the framework of fracture theory

[26].

Notably, this inconsistency parallels the flaw tolerance effect [28,29], where strength criteria [30] dominate fracture behavior at nanoscale asperities, overriding energy-based predictions. The flaw tolerance theory [28,29] might offer a plausible explanation for the observed correlation between friction and real contact area [15,17,18,20–25] provided elastic interactions between neighboring asperities are neglected. However, such interactions cannot be simply disregarded [31]. Studies reveal that only hierarchical, multiscale structures can effectively aggregate adhesive forces from nanoscale junctions [32], leaving unresolved the fundamental question: why is friction associated with contact area?

As a basic category of friction, the underlying mechanism of adhesive friction generally involves the cyclic formation and rupture of interfacial bonds. A molecular-scale stick-slip model [33] for rubber friction describes how polymer chains adhere to a moving surface, stretch, detach, and reattach. With subsequent refinements [34–36], the bell-shaped velocity dependence of friction with low-velocity side matching bulk viscoelastic temperature dependence was predicted [36], in agreement with experimental data [37,38]. Yet, these models appear to have overlooked the role of fracture mechanics and its size effect on adhesive friction. For example, rapid rupture fronts [39] were observed in friction and suggested to mediate the transition to frictional motion, which are well described by the fracture mechanics [40,41]. Current models of adhesive friction appear incomplete, lacking a unified framework to reconcile the continuum of distributed frictional dissipation with the discrete singularity of interfacial crack propagation [42].

Here, we investigate how the size effect in fracture theory can intriguingly vanish in adhesive friction when an elastic thin film slides against a rigid surface. With a simple adhesive frictional model accounting for the reattachment/healing of detached adhesive units, we then observe the emergence of multiple adhesive fronts with rich dynamics along the frictional interface. These adhesive fronts appear to mutually reinforce one another, leading to a macroscopic shear-off force that increases with contact area. We also observe that multiple fronts can undergo simultaneous detachment via rapid interfacial rupture. These findings are consistent with existing experiments, which may bridge the gap between tribology and fracture mechanics and be related to fault behaviors in earthquakes. This work is expected to provide fundamental insights into adhesive-friction coupling at interfaces with broad implications.

**Methods**

We examine a scenario where a generic 2-D elastic thin film with unit width, denoted as $w$, adheres to a smooth and flat rigid surface through multiple adhesive units distributed along the interface, as illustrated in Fig. 1. Though sophisticated adhesive elements were developed in the literature, we simplify each adhesive unit as a linear spring in our analysis without losing its generality. It should be emphasized that, while the general concept of adhesive friction governed by attachment/detachment mechanisms is well-established, its dependence on the size effect inherently governed by the fracture mechanics remains largely unexplored to our knowledge.

The elastic thin film has a length of $L$ and a height of $H$. It has an elastic modulus of $E$ and a Poisson's ratio of $\nu$. Adhesive springs with a spring constant of $k$ are uniformly distributed along the bottom surface of the elastic solid with a spacing of $d$, which are numbered sequentially from the left to the right. Adhesive sites along the rigid surface are also uniformly distributed with the same spacing of $d$. The elastic thin film is subjected to a horizontal sliding toward the right at a relatively low velocity of $V$.

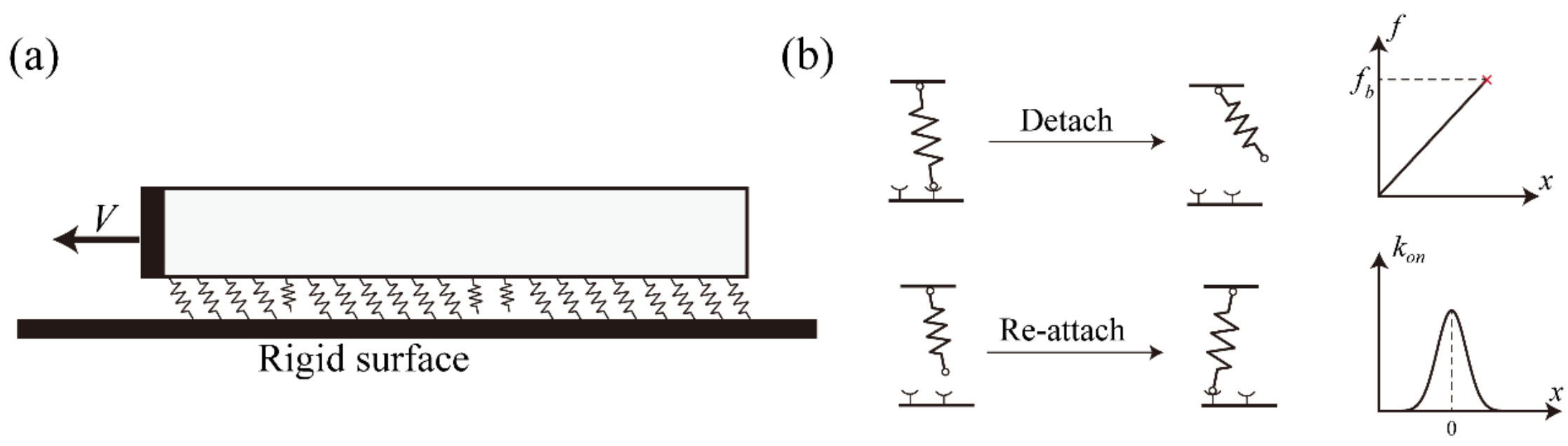

**Fig. 1.** (a) A generic mechanics model comprises an elastic thin film, adhering to a rigid surface through multiple adhesive units uniformly distributed along the interface. Without losing generality, each adhesive unit is simplified as a linear elastic spring. The elastic thin film is subjected to a horizontal sliding to the right with a velocity $V$. (b) In the model, an attached spring would deterministically detach at a critical force, $f_b$, and a detached spring can randomly re-attach to the interface to heal with a rate of $k_{on}$ guided by the elastic energy, both of which depend on the spring extension, $x$.

Initially, all adhesive springs attach to adhesive sites on the rigid surface with zero extension. The initial contact area, denoted as $A_0$, is given by $L$ times the unit width of a 2-D structure. When the elastic thin film is pulled, these springs would be stretched. A spring is assumed to detach from its adhesive site on the substrate once the force within it, denoted as *f*, reaches the adhesive strength of $f_b$. Detached springs are allowed to randomly re-attach to an open adhesive site to heal on the rigid surface at a rate of $k_{on}$, which is guided by the elastic energy. Without losing its generality, $k_{on}$ is assumed to follow the Kramer's law [43], given by $k_{on} = k_{on}^0 \cdot e^{\frac{-\Delta U}{k_B T}}$, where $\Delta U$ is the elastic energy that would be stored within the spring upon its potential attachment, $k_{on}^0$ is the attachment rate when $\Delta U = 0$, $k_B$ is the Boltzmann constant, and $T$ is the absolute temperature. As the elastic thin film slides, adhesive springs along the interface continue to detach and randomly re-attach to the rigid surface and the reaction force would be induced at the sliding end of the elastic solid, which would be taken as the adhesive frictional force.

The model is simulated with the coupled Finite Element Analysis and the Monte Carlo method. Within this numerical scheme, the deformation and the force within the model are calculated with Finite Element Analysis at each time step. If the force within an adhesive spring, denoted as *f*, reaches $f_b$, the spring would detach from the rigid surface. Otherwise, the Monte Carlo method will be employed to determine where and when a detached adhesive spring would randomly re-attach/heal to the rigid surface at the next time step. Default values of parameters used in the simulations are provided in Table 1.

**Table 1** Default values of parameters in the simulations

| Parameter | Value | Parameter | Value |
|---|---|---|---|
| $L$ | 1500 nm | $k$ | 0.3 pN/nm |
| $H$ | 100 nm | $k_{on}^{0}$ | 3 $s^{-1}$ |
| $E$ | 5 MPa | $d$ | 10 nm |
| $v$ | 0.3 | $V$ | 50 nm/s |
| $f_b$ | 10 pN | $k_B T$ | 4.14 pN.nm |
| $w$ | 1 nm | | |

**Results**

With the theoretical model described in Methods, we firstly investigate relatively thin elastic films sliding on a rigid surface with $H/d = 10$ and simulation results are displayed in Fig. 2. Without re-attachment of detached adhesive elements, i.e., $k_{on}^{0} = 0\mathrm{s}^{-1}$, we observe four distinct phases in the evolution of the frictional force, denoted as *F*. In Phase 1, the frictional force steadily increases, reaching a peak value in Phase 2, followed by a decrease to zero in Phase 3, and eventually maintaining at zero in Phase 4. The current adhered area, denoted as *A*, is calculated by multiplying the number of attached adhesive springs by $d$ and unit width, which gradually diminishes to zero within a relatively short shear displacement in Fig. 2b. The prediction from the Kendall model [27], which determines the peeling strength of a thin elastic film being adhered to a rigid substrate, denoted as $F_{0T}$, is given by be $f_b\sqrt{\frac{EHw}{kd}}$, which yields a value of 130 pN, aligning closely with the peak value of the frictional force in Phase 2 obtained in our numerical simulation.

For non-zero but relatively small $k_{on}^{0}$, Fig. 2a shows notable differences arising in Phase 2 and Phase 4. In Phase 2, the frictional force exhibits an increase with shear displacement that surpasses the prediction of Kendall's theory [27]. In Phase 4, the frictional force is no longer zero and exhibits fluctuation, which closely resembles the stick-slip phenomenon with approximate periodicity, where it generally increases during the "stick" phase and subsequently abruptly decreases during the "slip" phase. Notably, the peak frictional force in Phase 4 is significantly lower than that in Phase 2. Figure 2b reveals that a portion of adhesive springs attach to the substrate with its number fluctuating in Phase 4. As $k_{on}^{0}$ increases, the peak frictional force both in Phase 2 and Phase 4 increases. With further increase in $k_{on}^{0}$, Phase 2 becomes indistinguishable and the frictional force in Phase 4 displays significant fluctuations. Importantly, during these fluctuations, the peak frictional force can exceed by a large margin the prediction provided by the classical Kendall model [27]. For instance, when $k_{on}^{0}d/V = 0.6$, the former can surpass the latter by more than 200%. In Fig. 2b, it is evident that a large portion of springs attach to the substrate, although their number fluctuates. When plotting out the largest peak frictional force in Phase 4, denoted as $F_p$, against $k_{on}^{0}$ in Fig. 2c, we find that it generally increases with $k_{on}^{0}$ until it saturates at a large $k_{on}^{0}$.

To see how the peak frictional force in Phase 4 with large $k_{on}^{0}$ can be even much larger than the prediction made by the classical Kendall's model [27], we plot out representative adhesive forces along the interface in Fig. S1. When $k_{on}^{0}$ = 0, only a single region of crack-like force profile, i.e., the adhesive front, is observed along the interface, where only a small number of adhesive springs are strained, while the remaining adhesive springs exert no force. In contrast, with a large $k_{on}^{0}$ value, such as $k_{on}^{0}d/V = 0.6$, at least two regions of crack-like force profiles, i.e., two adhesive fronts, may emerge in close proximity along the interface, as shown in Fig. S1c. Consequently, a significant portion of adhesive springs would experience strain along the interface, resulting in a much higher frictional force than predicted by Kendall's model [27], despite non-uniform forces within the adhesive springs.

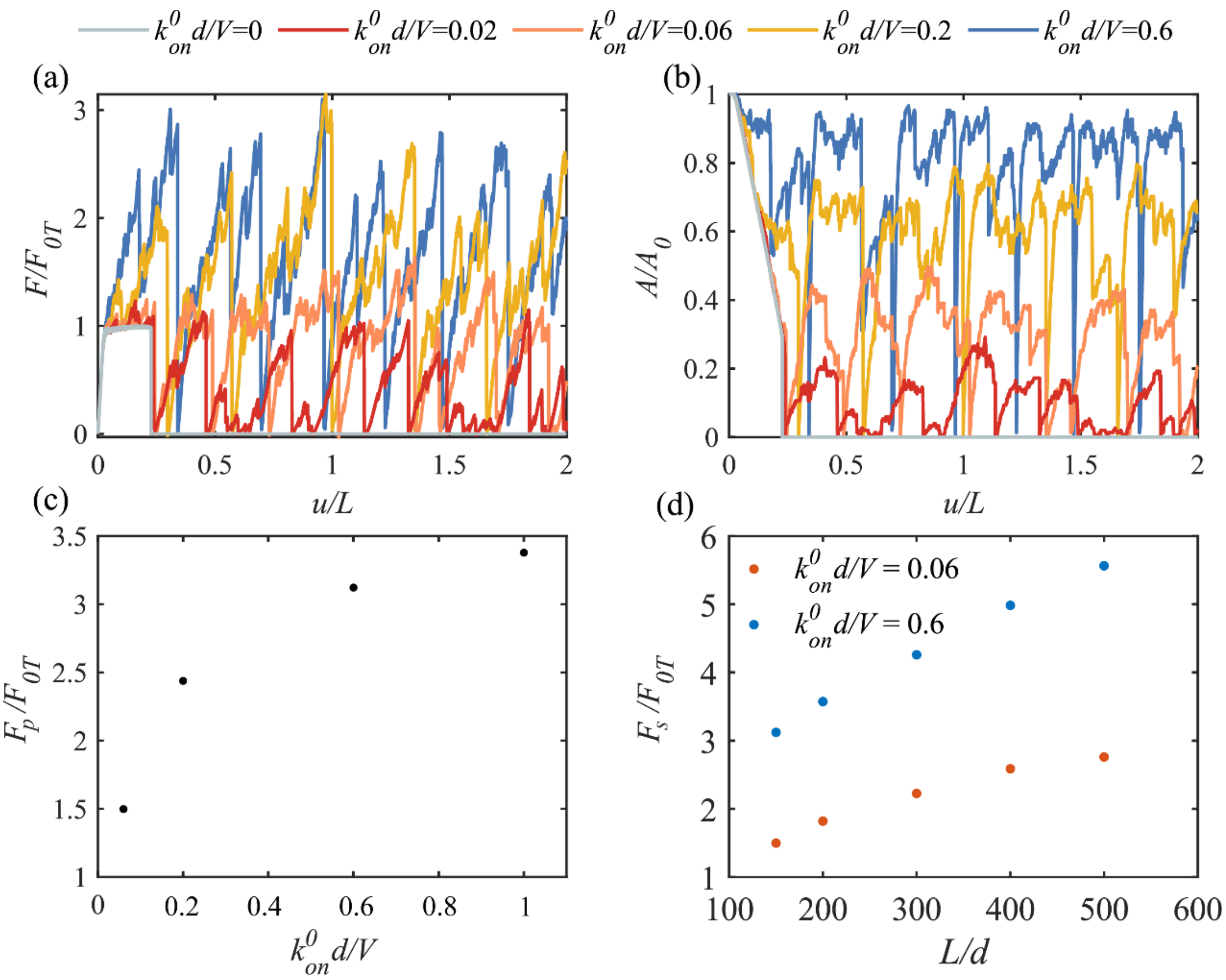

**Fig. 2.** Simulation results for relatively thin elastic films: Variation of the frictional force (a) and the current adhered area (b) with the sliding displacement at different $k_{on}^{0}$; (c) Effect of $k_{on}^{0}$ on the peak frictional force in Phase 4; (d) The variation of the shear-

off force with $L$. In the analysis, $H$ = 10$d$.

The adhesive friction was observed to exhibit a linear relationship with the real contact area in experimental studies [44,45]. To see if our model can generate comparable predictions, we adjust the length of the elastic solids in our simulations. In the analysis, we calculate the average of the peak frictional forces in Phase 4, termed as the shear-off force, denoted as $F_s$. To calculate the shear-off force in Phase 4, we run simulations of pulling thin film for 100 trajectories and also ensure the occurrence of at least one stick-slip event within each trajectory. In Fig. 2d, we observe that $F_s$ appears to increase with $L$. To gain further insight, we plot particular adhesive forces and also the corresponding longitudinal strains along the interface at certain sliding displacements in Figs. 3a-f. These figures reveal that the number of adhesive fronts generally increases with $L$. It appears that these adhesive fronts effectively support each other, leading to the persistent increase in the shear-off force as the interface area increases. If the friction force generated within each adhesive front is assumed to be $F_{0T}$, the size of each adhesive front equal and no gap exists between neighboring adhesive fronts for $k_{on}^{0}d/V = 0.6$, then the size of each adhesive front is estimated to be ~140d, which is comparable to the simulated size of the single adhesive front without re-attachment displayed in Fig. S1. Comparing Figs. 3a-c with Figs. 3d-f, it can also be observed that a lower reattachment rate can not only result in a reduced density of adhesive fronts along the interface but also decrease the frictional force induced within individual adhesive fronts.

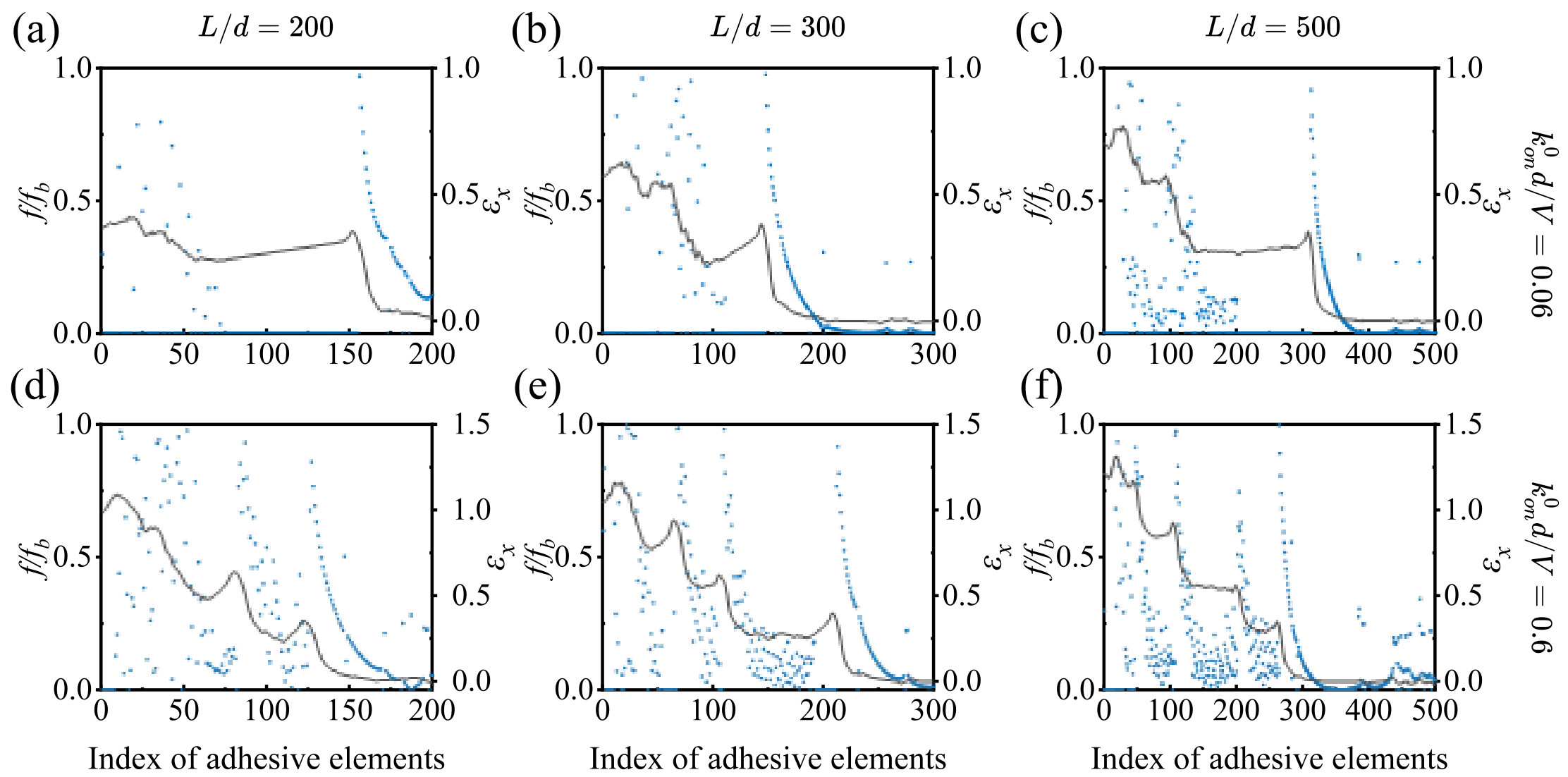

**Fig. 3.** Effects of $L$ of relatively thin elastic films on the adhesive friction: The particular distribution of adhesive forces, represented with dots, and the corresponding longitudinal strains of $\varepsilon_x$, represented with curves, along the interface at certain sliding displacements when $k_{on}^0 d/V = 0.06$ (a-c) or $k_{on}^0 d/V = 0.6$ (d-f). In the analysis, $H = 10d$.

The cohesive zone size in relatively thin elastic films can be significantly constrained by film thickness [31], when the thickness is substantially smaller than the cohesive zone size in bulk solids. We then increase the thickness of the thin films in our analysis and furtherly investigate relatively thick elastic films with $H/d = 100$, where $H$ would be comparable to the cohesive zone size in bulk solids and the elastic fields would be in a relatively complex stress state. The frictional force during Phase 2, denoted as $F_{0B}$, slightly varies in Fig. 4a when $k_{on}^0 = 0$, due to the resulting minor bending along the interface. In Fig. S2, only a single adhesive front exists along the interface when $k_{on}^0 = 0$. For non-zero $k_{on}^0$, the frictional force in Phase 4 displays the stick-slip phenomenon with approximate periodicity. Figure 4b reveals that a portion of adhesive springs attach to the substrate with its number fluctuating in Phase 4. In Fig. 4c, the peak frictional force, $F_p$, in Phase 4 shows a rising trend with $k_{on}^0$ until it reaches a saturation point at large $k_{on}^0$, which can be notably higher than the case without re-attachment. Additionally, in Fig. S2, it can be observed that a significant proportion of adhesive springs along the interface experience strain for large values of $k_{on}^0$. When varying $L$ in our simulation, we find that the shear-off force persistently increases with $L$ in Fig. 4d. Figures. 5a-f and also Fig. S3 reveal that the number of adhesive fronts generally increases with $L$ and the separation between neighboring adhesive fronts can be smaller than the size of an adhesive front, which is in turn comparable to the simulated size of the single adhesive front without re-attachment displayed in Fig. S2. These findings on relatively thick elastic films align closely with the observations made in relatively thin elastic solids, suggesting their general applicability to elastic thin films.

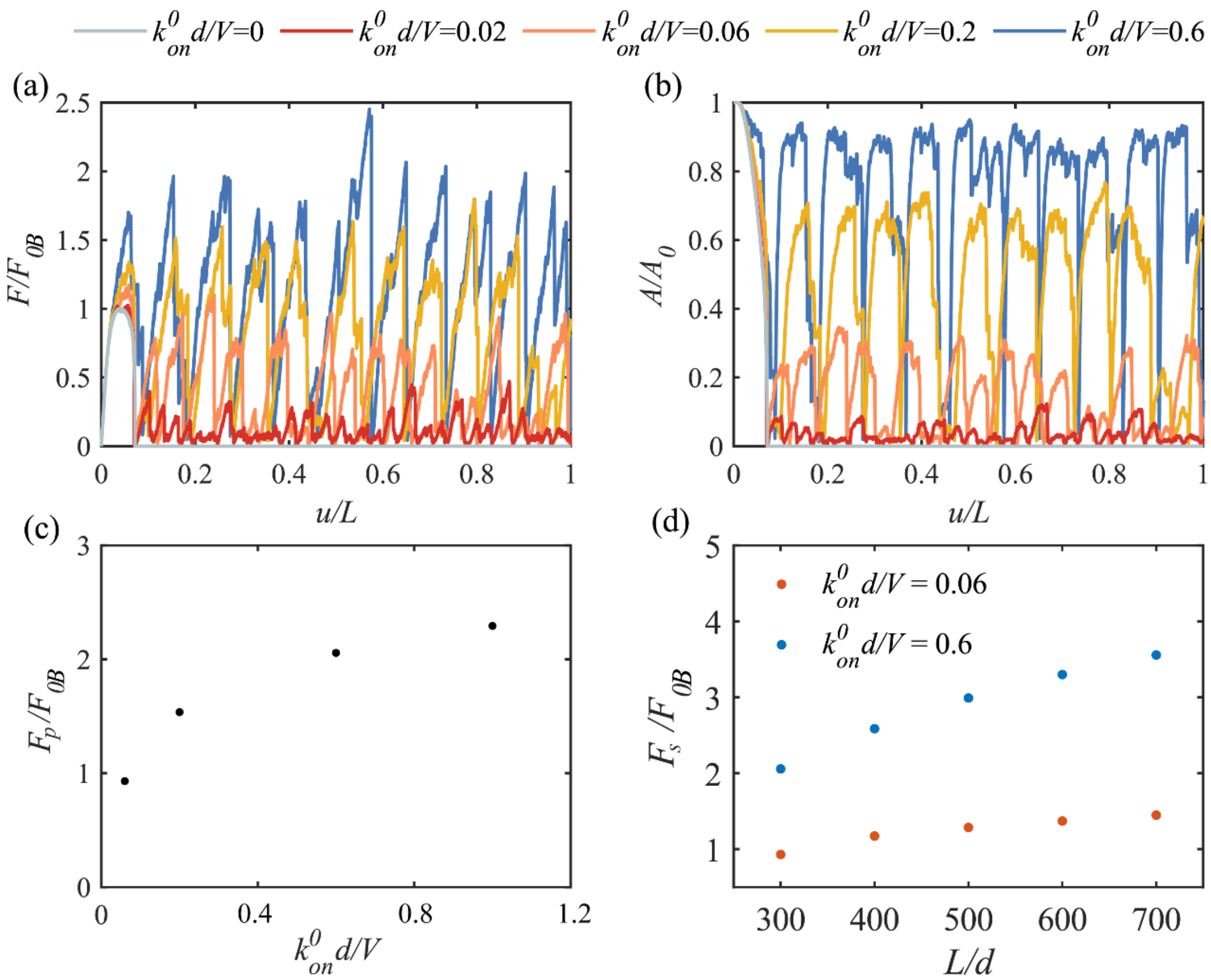

**Fig. 4.** Simulation results for relatively thick elastic films: Variation of the frictional force (a) and the current adhered area (b) with the sliding displacement at different $k_{on}^{0}$; (c) Effect of $k_{on}^{0}$ on the peak value of the frictional force in Phase 4; (d) The variation of the shear-off force with $L$. In the analysis, $L = 300d$ for (a-c) and $H = 100d$.

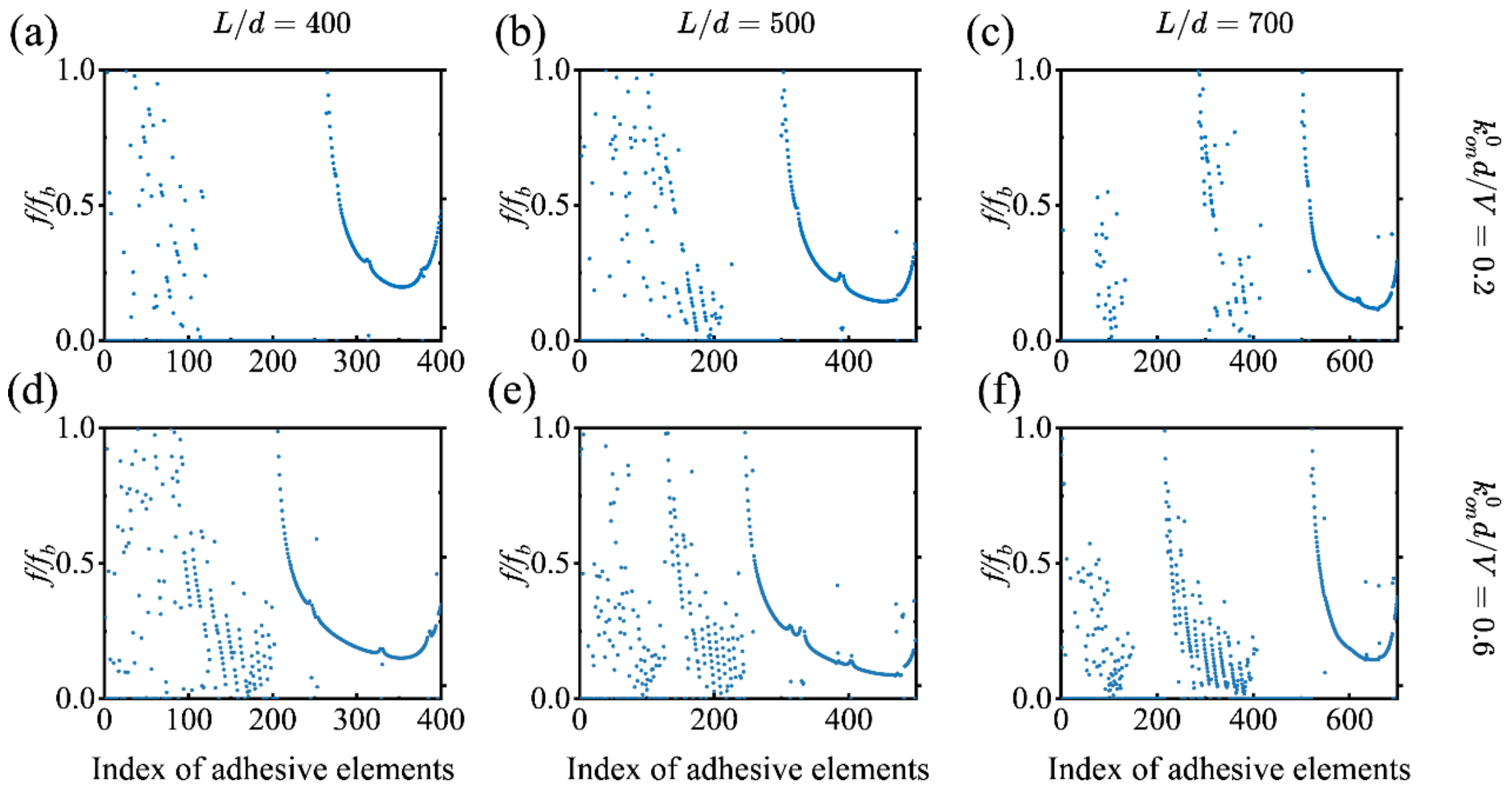

**Fig. 5.** Effect of $L$ of relatively thick elastic films on the adhesive friction: The particular distribution of adhesive forces along the interface at certain sliding displacements when $k_{on}^0 d/V = 0.2$ (a-c) or $k_{on}^0 d/V = 0.6$ (d-f). In the analysis, $H = 100d$.

To provide more insights into the mechanisms of the adhesive friction, we plot out typical time evolution of adhesive fronts along the interface. As displayed in Fig. 6, adhesive fronts in our model analysis appear to initiate at the loading edge and propagate outward. Healing of broken adhesive units can disrupt stress decay pattern and create non-monotonic spatial profiles prominently within the front closest to the loading edge. During propagation, individual fronts exhibit repeated gradual advancement and instability-driven jumps. Pronounced drops in frictional force correlate with dominant front instabilities, which are followed by the formation of new adhesive fronts within the reformed contact zone to drive frictional force recovery. Strikingly, the formation of new adhesive fronts locally occurs in an apparently ordered, sequentially retrograde manner. The relative positioning of neighboring adhesive fronts evolves dynamically. It appears that the elastic coupling between neighboring fronts induces resistance, leading to their coordinated propagation or occasionally merging, and neighboring fronts frequently simultaneously lose stability. The self-replenishing process continues until front-front interactions drives the system possibly to a critical packing density, precipitating another dominant front instability. The largest force drop coincides with synchronized failure across all fronts along the interface. These high-resolution observations demonstrate that the spatiotemporal evolution of emergent multiple adhesive fronts along the interface follows self-organized critical dynamics in response to mechanical loading.

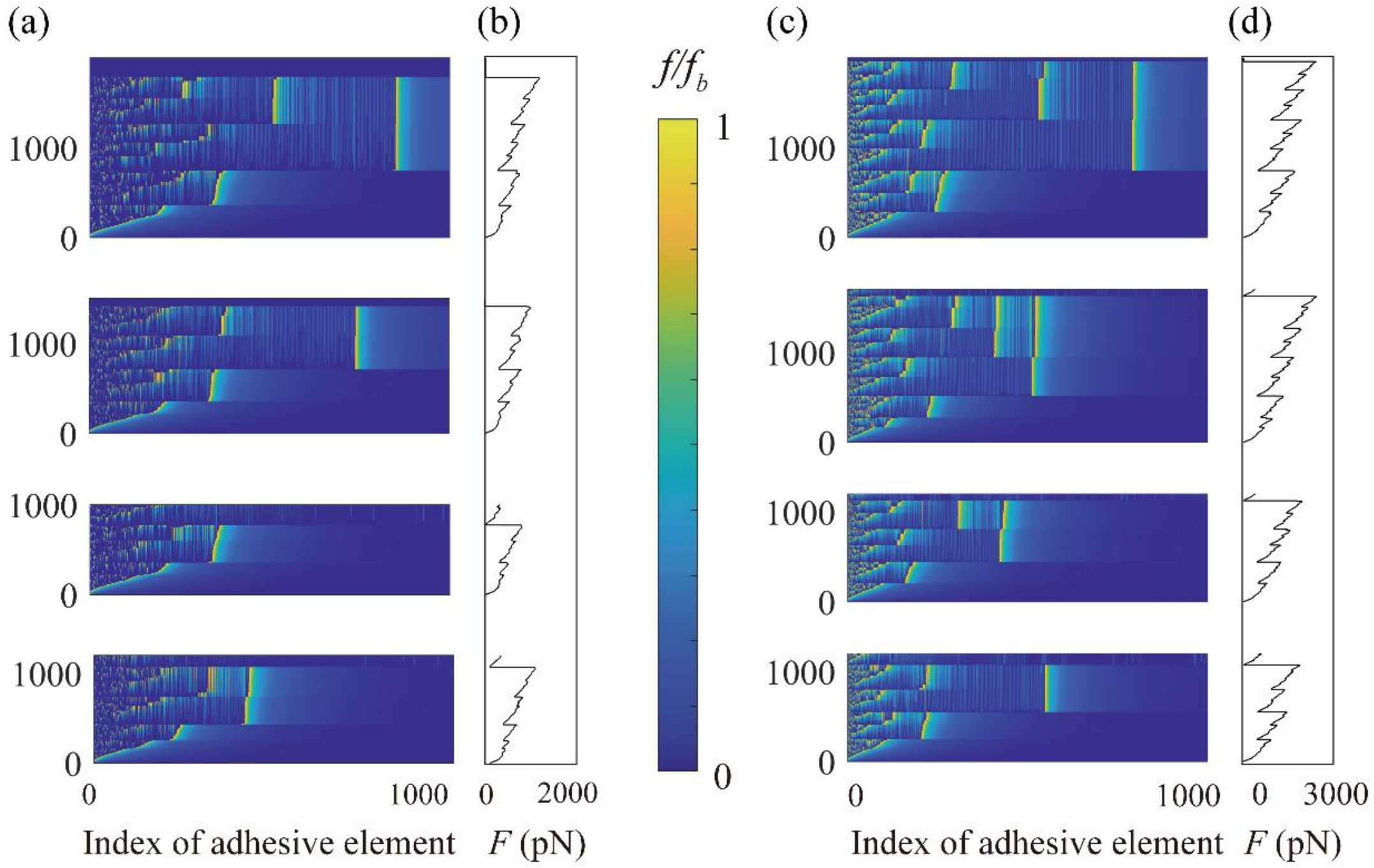

**Fig. 6.** Contour plots show four time evolutions of adhesive fronts along the interface with low re-attachment rate ($k_{on}^0 d/V = 0.06$) (a) and with high re-attachment rate ($k_{on}^0 d/V = 0.6$ ) (c) together with respective evolution of corresponding frictional forces (b, d). In the simulation, $L = 1000d$.

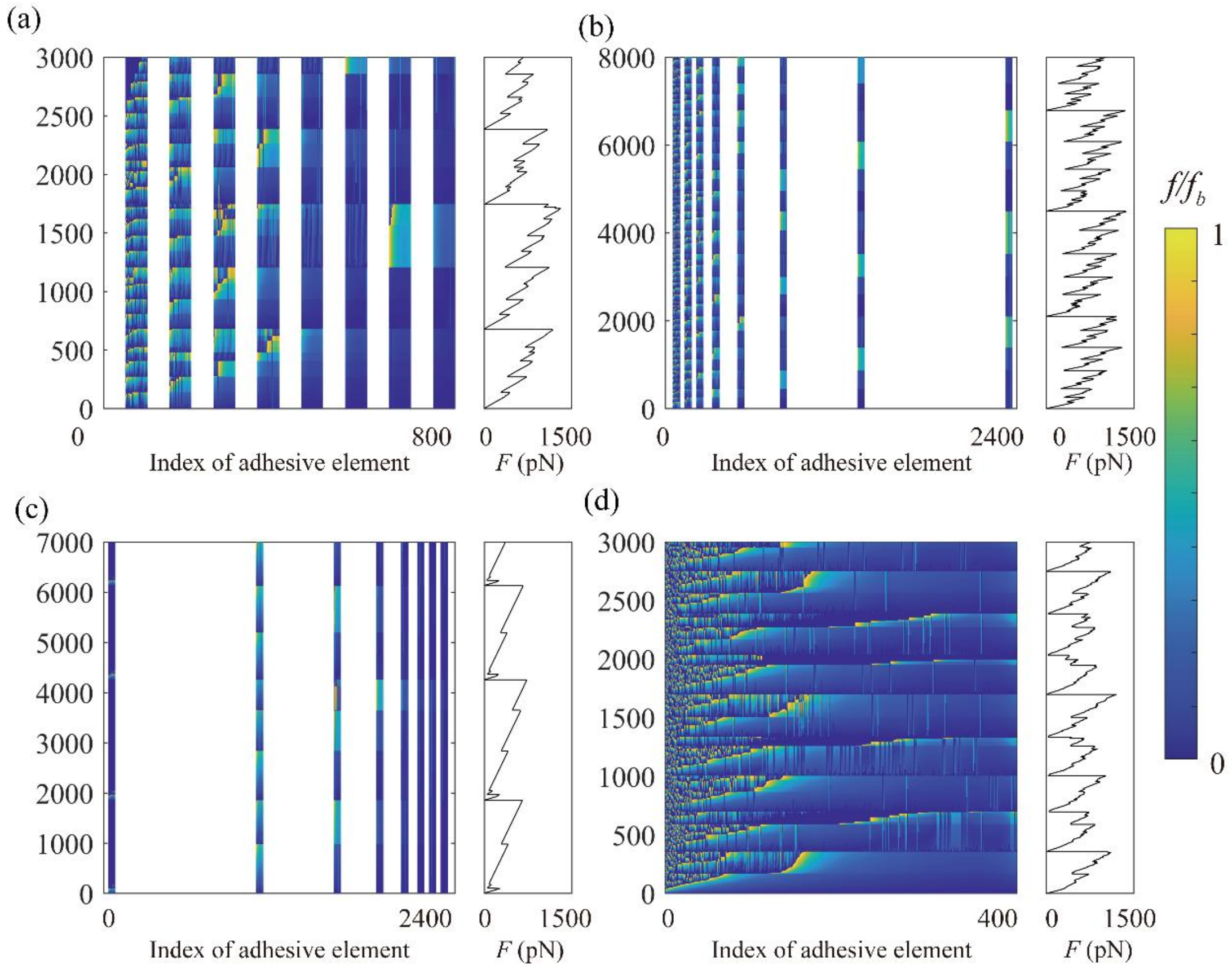

**Fig. 7**. Effects of adhesive defects on the adhesive friction: Contour plots show typical time evolution of adhesive fronts along the interface together with the respective time evolution of corresponding frictional forces with uniform spacing between neighboring asperities (a), gradually increasing spacing between neighboring asperities (b), and gradually decreasing spacing between neighboring asperities (c). For comparison, results for an interface without adhesive defects are also plotted out in (d). In the analysis, $k_{on}^{0}d/V = 0.6$ and $L = 400d$.

Since asperities are naturally present along most interfaces, we investigate their influence on adhesive friction by considering three distinct distributions of adhesive defects in our analysis. The total contact area of the thin films in the analysis is fixed at 400*d*, with each individual asperity occupying 50*d*. In the first configuration, asperities are uniformly spaced with a constant interval of 50*d*, termed as case 1. The second configuration features a gradually increasing spacing between asperities from left to right (30*d*, 30*d*, 60*d*, 120*d*, 240*d*, 480*d*, 960*d*), termed as case 2, while the third configuration exhibits the inverse pattern, with spacing gradually decreasing from left to right (960*d*, 480*d*, 240*d*, 120*d*, 60*d*, 30*d*, 30*d*), termed as case 3. Figures 7a-c illustrate the evolution of adhesive fronts for each interface type, along with their corresponding frictional force profiles. For comparison, Fig. 7d presents the results for an interface without adhesive defects as case 4. Comparative analysis of Fig. 7 reveals several key findings: (1) the shear-off force in case 3 is significantly lower than those in the other cases; (2) the frictional forces along some asperities are almost uniform, resembling the flaw tolerance effect [28,29] and the frictional force distribution in case 2 (increasing spacing) is substantially broader; and (3) the front evolution patterns in cases 2 demonstrate greater regularity. These results collectively demonstrate that asperity distribution can profoundly affect the magnitude and spatial characteristics of frictional adhesion and also the dynamics of adhesive fronts.

**Discussions**

Using a simplified adhesive frictional model, we demonstrate that re-attachment /healing processes at the molecular level generate multiple discrete adhesive fronts along the interface. These fronts, comparable in size to the cohesive zone described in classical fracture mechanics, exhibit rich dynamics and can mutually reinforce one another, leading to the increase in adhesive friction with interface area. This finding is consistent with experiment [15,17–25] and explains why friction is associated with contact area, but defies classical fracture theory predictions [26].

Our analysis reveals that frictional forces display pronounced spatial heterogeneity at the scale of the cohesive zone along the interface and indicates that when the applied shear force remains below the critical shear-off threshold, stress is gradually released through a sequence of discrete adhesive front events. This observation aligns with prior experimental studies [46], where shear loading triggered crack-like precursor activity along frictional interfaces, dynamically redistributing contact stresses. Indeed, while the motion of rapidly propagating ruptures at frictional interfaces [19,47] can resemble shear cracks described by fracture mechanics [48,49], their nucleation process within the precursor zone still remains poorly understood [50], which, however, is revealed in our analysis. Our analysis shows that the rapid rupture is preceded by the dynamic piling up process of slow fronts. As more and more slow fronts are piled up possibly to a critical packing density, indicated from Fig. 6, our analysis suggests that the leading front's propagation transitions to a state governed by the classical fracture mechanics, dictating the subsequent system wide crack propagation process.

Previously, a linear scaling relationship between the so-called precursor length along a frictional interface, denoted as $l_p$, and the applied shear force, denoted as $F_a$, was proposed based on the experimental observation [46]. Our analysis furtherly suggests that the linear relationship between $l_p$ and $F_a$ arises from cooperative effects of densely packed adhesive fronts within the precursor zone, where each front may contribute a quantized friction increment, such that

$$F_a \propto \rho \Delta f l_p \tag{1}$$

where $\rho$ is the density of the adhesive fronts piling up along the interface within the precursor zone, $\Delta f$ is the average frictional force generated within a single adhesive front, both of which are expected to depend on the re-attachment/healing rate and the sliding velocities. In order to align with Amonton's law, $\rho \Delta f l_p$ in Eq. (1) is expected to be proportional to the normal load.

In our analysis, inertial effects are entirely neglected—an assumption justified only under relatively low-velocity conditions. However, it is conceivable that the progressive accumulation of adhesive fronts could induce a substantial increase in the rupture rate of the individual fronts. To estimate the propagation speed of rupture fronts during catastrophic detachment triggered by closely spaced multiple adhesive fronts, we employ a spring-block model connected to a thin film (see Appendix), as illustrated in Fig. A1. As seen from Figs. A2a-c, our analysis then suggests that the rupture velocity can indeed approach the Rayleigh wave speed as the number of adhesive fronts increases. Concurrently, Fig. A2d reveals the growth in rupture propagation distance with increasing number of adhesive fronts. These observations imply the potential relevance of our findings in this work to seismic phenomena, where earthquake rupture along frictional interfaces between tectonic boundaries can propagate at a very high speed [48,51].

Fault healing was proposed as a critical mechanism for both slow earthquakes and, surprisingly, catastrophic earthquakes [52,53]. Slow slips along frictional interfaces between tectonic boundaries are regarded to represent a fundamental process in fault dynamics [52,54–56]. Yet, the mechanistic links connecting slow slips, slow earthquakes, and catastrophic earthquakes are still not fully resolved. The rate-and-state friction law [57,58], which characterizes the rate- and state-dependent evolution of frictional resistance, has been widely employed to model earthquake-related fault dynamics. Theoretical studies incorporating the velocity-weakening rate-and-state friction law have successfully reproduced slip pulses [52,54,59–61]. In our simple model, we explicitly account for the detachment and subsequent reattachment/healing of individual adhesive elements. This approach also yields velocity-weakening frictional behavior. Interestingly, our simulations reveal that slow slips frequently occur during the stick phase, which are analogous to slow slips, slip pulses, or slow earthquakes in fault dynamics. Meanwhile, the slip phase in our analysis culminates in rapid energy release, analogous to catastrophic earthquakes. Our analysis clearly shows how multiple adhesive fronts emerge and closely pile up along the frictional interface, where slow slips may propagate, arrest, or ultimately trigger a dynamic transition to a catastrophic rupture. Thus, our study in this work may fundamentally connect slow slips, slow earthquakes and catastrophic earthquakes, revealing a unified framework for fault behavior.

Adhesive fronts were also reported in previous theoretical works [62,63], which provided important insights into understanding the mechanisms in friction. It was demonstrated through multiscale spring-block modeling [63] that slow fronts exhibit different characteristics from fast ones, with the transition between regimes arising from slow slip processes. However, their observations [63] were limited to only a couple of adhesive fronts with little information being provided for the dynamics of multiple piling up adhesive fronts. With rate-and-state frictional laws incorporating aging effects, creep patches analogous to adhesive fronts were also identified [62], which may undergo linear instability at critical nucleation sizes - a phenomenon governed by the competition between frictional weakening and size-dependent bulk stiffness [64]. In comparison, with a simple adhesive frictional model, our current investigation reveals rich behaviors of adhesive fronts emerging from reattachment/healing at the molecular level. We observe multiple adhesive fronts with characteristic dimensions comparable to the cohesive zone size undergoing complex spatiotemporal interactions. Crucially, we demonstrate that this self-organized pattern formation enables fracture-mediated processes to emulate area-proportional friction - thereby resolving the long-standing conceptual divide between fracture mechanics and tribological frameworks and providing the fundamental insight into the origin of macroscopic frictional forces' association with contact area.

Finally, we must acknowledge that, while our current simplified model focuses on revealing fundamental adhesive friction mechanisms and provides an explaination for how the frictional force is associated with contact area, as reported in various experiments [15,17–25], future refinements of our model may incorporate rate-dependent bond strength, time-dependent creep behavior, stochastic asperity statistics, etc., to enable quantitative predictions to be directly compared with experiments of specific frictional systems. Critical model parameters in the future analysis can also be experimentally determined, for example, through functionalized AFM pull-off tests for bond strength [65], Surface Forces Apparatus studies of rebinding kinetics under confinement [66], and AFM peak-force tapping mode for topographic analysis[67].

## Conclusion

In this work, we investigate adhesive frictional forces along an interface capable of re-attachment/healing. In our analysis, we identify multiple adhesive fronts exhibiting with rich dynamics that accumulate along the interface. The number of these fronts generally increases with interface area, which can lead to the association of the shear-off force with precursor dimensions. Based on these findings, we propose a refined law of adhesive friction. Our analysis also reveals that the piling up of multiple adhesive fronts along the interface can drive individual fronts to propagate like a crack at a high speed, which potentially bridges the gap between tribology and fracture mechanics. The relevance of our work to earthquakes is discussed. We expect that the implications of our findings in this work can extend across diverse fields where healing governs interfacial behavior.

## Acknowledgments

This work was supported by the National Natural Science Foundation of China (Grant No.: 12372318, 11872334), and Zhejiang Provincial Natural Science Foundation of China (Grant No.: LZ23A020004).

**Author contributions:** B.C. designed research, P.C. and B.C. performed research, P.C. and M.Y. analyzed data, P.C. and B.C. wrote the manuscript.

**Competing Interest Statement:** The authors declare no competing interests.

## Appendix

As illustrated in Fig. A1, a portion of thin film adhered through discrete adhesive fronts is simplified with rigid blocks connected through elastic springs. The total number of blocks or springs is denoted as *N*. Right before the occurrence of catastrophic detachment of *N* blocks, the adhesion force on each block and also on the rest of thin film is assumed to be the same, denoted as $F_0$. Based on force equilibrium, the force within the $i^{\text{th}}$ spring is then given by $F_i = (N - i + 1)F_0$. The spring constant of each spring is considered to be in inverse proportion to the critical spacing between neighboring blocks at this moment, which is assumed to be same and denoted as *l*. The spring constant of all springs is then the same and denoted as $k$. The total elastic stretch of all springs would be given by $\sum_{i=1}^{N} \frac{(N-i+1)F_0}{k}$.

Consider that the catastrophic detachment of *N* blocks then occurs under the critical condition. Neglecting the inertia effect of blocks, the force within each spring would become the same. This force can drive the leading crack front along the interface to propagate forward at a velocity of $v_p$. To estimate $v_p$, we neglect the effect of loading velocity of the thin film, which can be much smaller than $v_p$. Considering that the leading crack front moves forward a distance of *s*, the instant peeling force on thin film, denoted as $F_p$, will be given by

$$F_p = \frac{N(N+1)F_0}{2(N + s/l)}, \tag{A1}$$

which decreases as *s* increases. The corresponding stress intensity factor for a static crack, denoted as $K_{II}(0)$, is given by [68]

$$K_{II}(0) = \frac{F_p}{w\sqrt{2H}}. \tag{A2}$$

Based on the fracture criterion for a propagating crack [26],

$$K_{II}(0)k(v_p) = K_{IIC}, \tag{A3}$$

where $k(v_p)$ is a geometry independent function of crack propagation speed and can be approximately given by $k(v_p) = 1 - v_p/c_R$ [26], with $c_R$ being the Rayleigh wave speed, and $k_{IIc}$ represents the critical stress intensity factor resisting the material to dynamic crack propagation. With Eqs. (A1-A3), $v_p$ is derived to be

$$v_p = c_R\left(1 - \frac{2(N + s/l)}{N(N+1)}\frac{K_{IIc}w\sqrt{2H}}{F_0}\right). \tag{A4}$$

With Eq. (A4), the variation of $v_p$ with *s* for different *N* or $K_{IIC}\sqrt{H}/F_0$ is plotted out in Figs. A2a-c by noting that *w* corresponds to unit width. The critical crack propagation distance, denoted as $s_c$, upon which $v_p$ approaches 0 is also plotted out for potential interest.

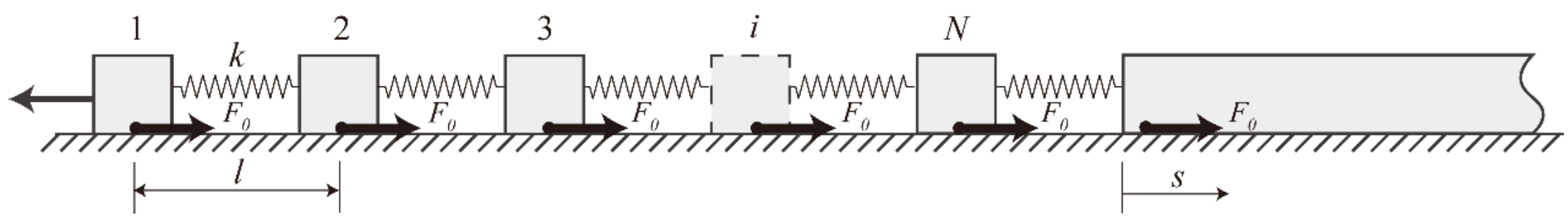

**Fig. A1.** The portion of thin film with discrete adhesive fronts being piled up under loading is simplified with rigid blocks connected through elastic springs. The total number of blocks or springs is *N*. Right before the occurrence of catastrophic detachment of *N* blocks, the adhesion force on each block and also on the rest of thin film is $F_0$ and the critical spacing between neighboring blocks is *l*.

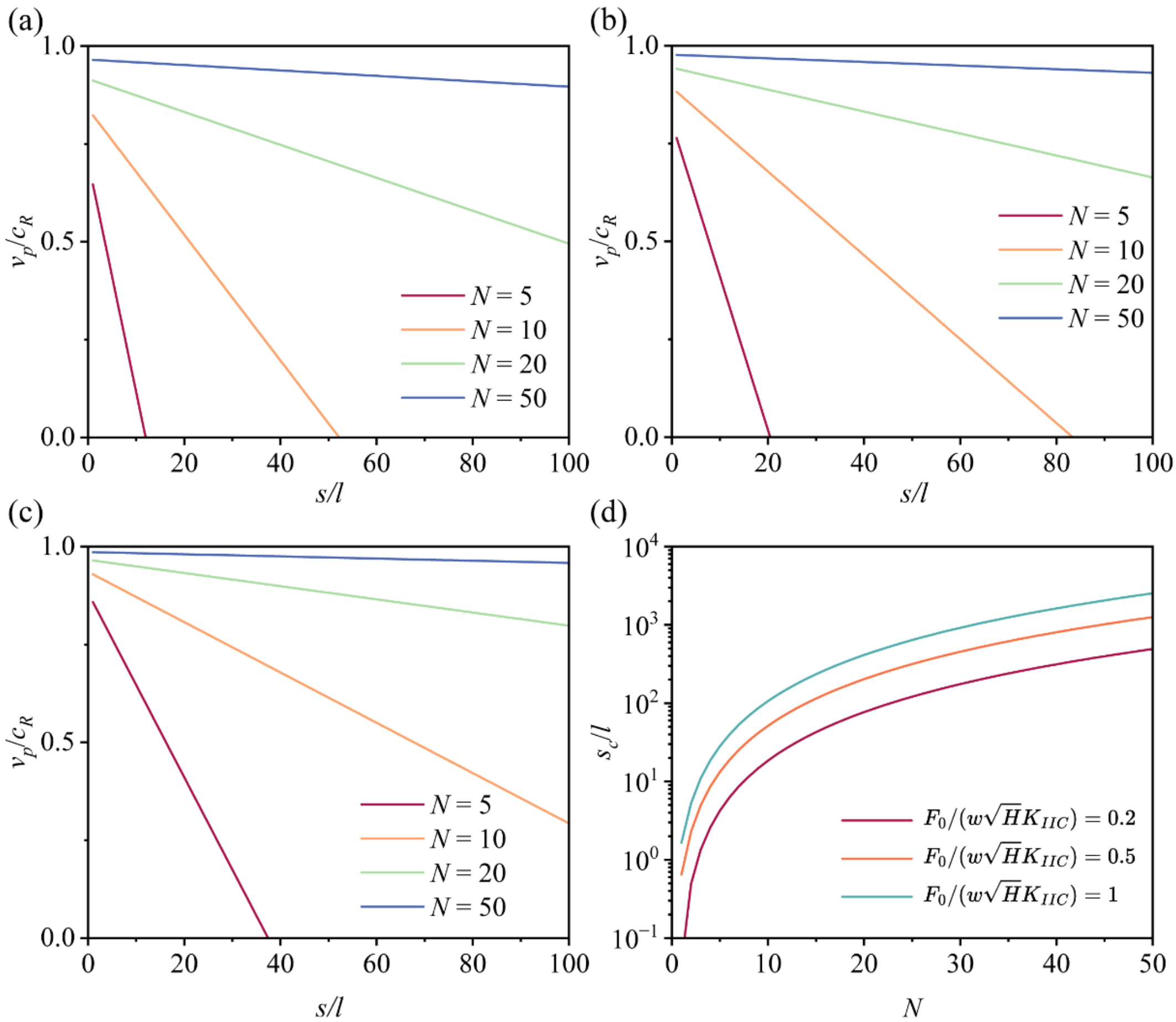

**Fig. A2.** Variation of $v_p$ with $s$ when $F_0/(w\sqrt{H}K_{IIC})$ is equal to 0.2 (a), 0.4 (b), and 0.5 (c), respectively. (d) Increase of the number of blocks leads to the increase of $s_c$.

## Supplementary Information

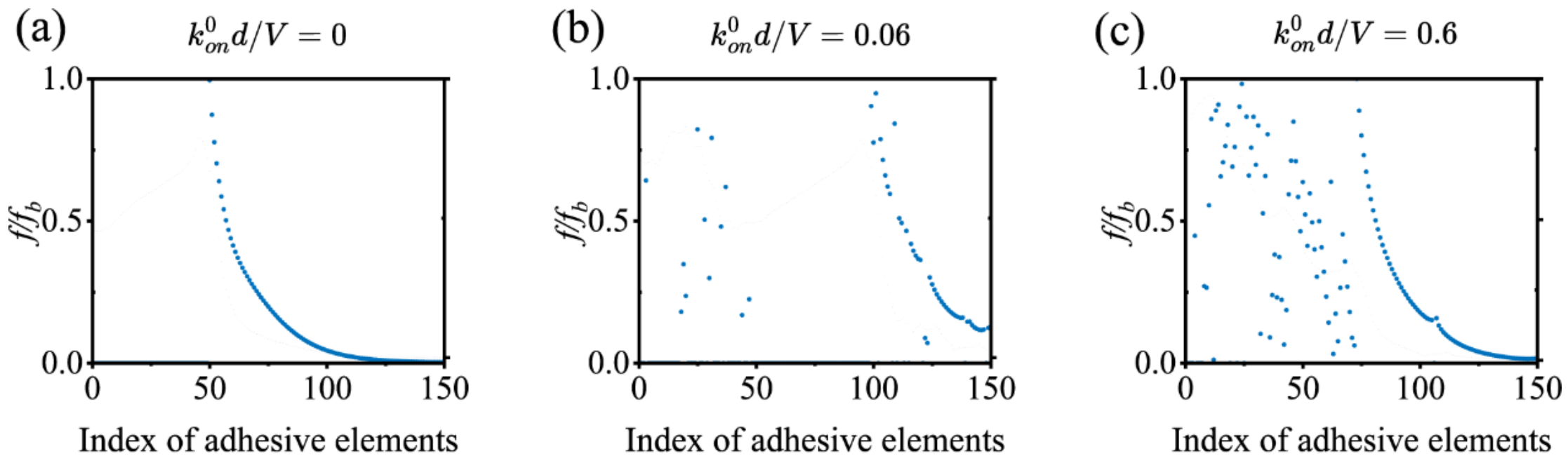

**Figure S1.** The particular distribution of adhesive forces along the interface of relatively thin elastic films when $k_{on}^0 d/V = 0$ (a), and $k_{on}^0 d/V = 0.06$ (b) and $k_{on}^0 d/V = 0.6$ (c). In the analysis, $H = 10d$.

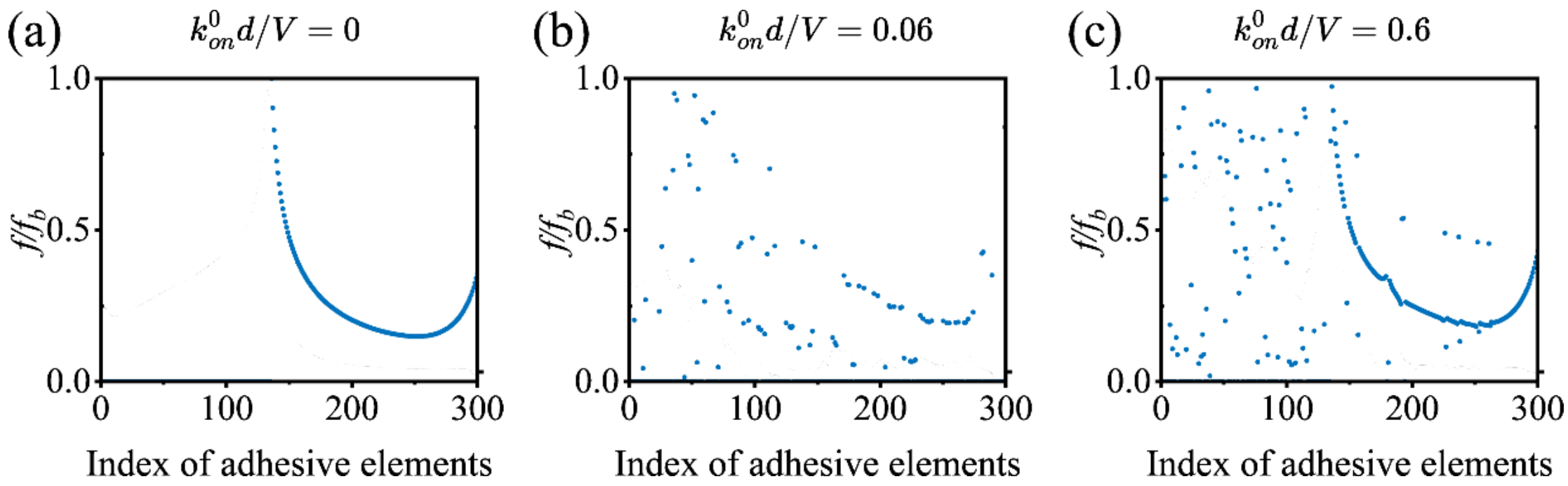

**Figure S2.** The particular distribution of adhesive forces along the interface of relatively thick elastic films when $k_{on}^0 d/V = 0$ (a), and $k_{on}^0 d/V = 0.06$ (b) and $k_{on}^0 d/V = 0.6$ (c). In the analysis, $H = 100d$.

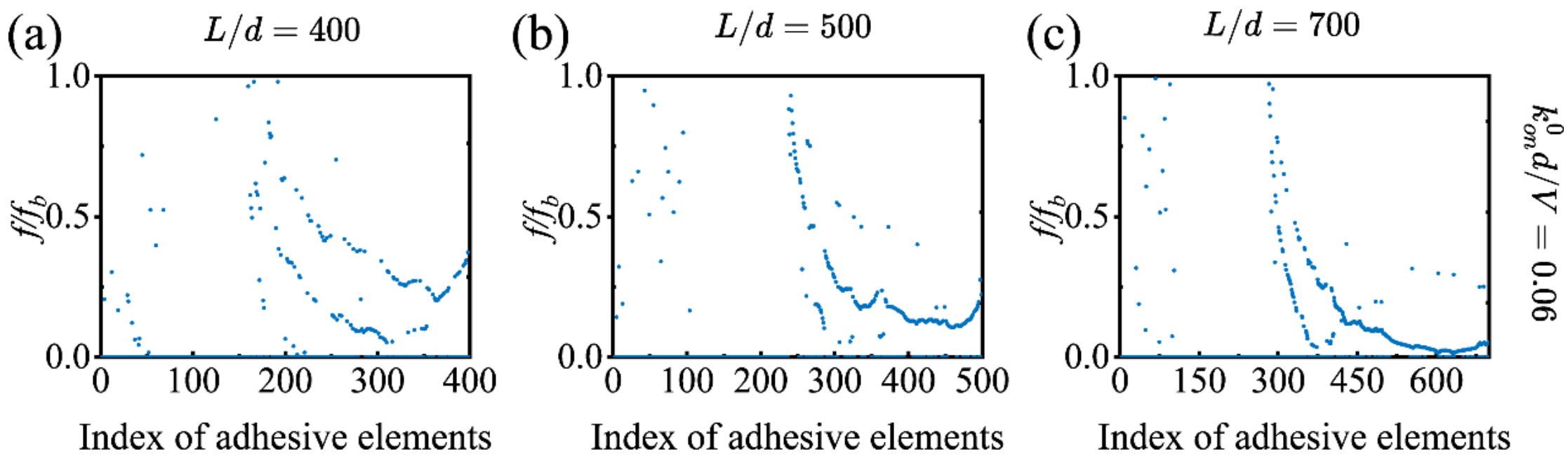

**Figure S3.** Effect of $L$ of relatively thick elastic films on the adhesive friction: The particular distribution of adhesive forces, represented with dots, along the interface for different $L$ when $k_{on}^{0}d/V = 0.06$. In the analysis, $H = 100d$